\documentclass[sigconf,authorversion]{acmart}
\renewcommand\footnotetextcopyrightpermission[1]{} 
\usepackage{colortbl}
\usepackage{xcolor}
\PassOptionsToPackage{table, xcdraw}{xcolor} 

\usepackage{multirow}
\usepackage{makecell}
\usepackage{tcolorbox}

\definecolor{warningcolor}{RGB}{255,97,0}
\definecolor{Morality}{HTML}{b6e3e7} 
\definecolor{Harm}{HTML}{fbf1d7}  
\definecolor{Misinformation}{HTML}{afe1af} 
\definecolor{Offensive}{HTML}{fad6b5}         
\definecolor{Discrimination}{HTML}{faadac}        
\definecolor{Privacy}{HTML}{fcdfe5}        
\definecolor{Illegal}{HTML}{daf1ee}           

\tcbuselibrary{skins, breakable, theorems}

\AtBeginDocument{%
  }

\setcopyright{acmlicensed}
\copyrightyear{2018}
\acmYear{2018}
\acmDOI{XXXXXXX.XXXXXXX}
\acmConference[Conference acronym 'MM]{Make sure to enter the correct
  conference title from your rights confirmation email}{
October 27--31,
  2025}{Dublin, Ireland}
\acmISBN{978-1-4503-XXXX-X/2018/06}

\begin{document}

\title{ShieldVLM: Safeguarding the Multimodal Implicit Toxicity via Deliberative Reasoning with LVLMs}


\author{Shiyao Cui}
\authornote{Both authors contributed equally to this research.}
\email{cuishiyao@mail.tsinghua.edu.cn}

\author{Qinglin Zhang}
\authornotemark[1]
\affiliation{%
  \institution{The Conversational AI (CoAI) group, DCST, Tsinghua University}
  \country{China}
}

\author{Xuan Ouyang}

\author{Renmiao Chen}
\affiliation{%
  \institution{The Conversational AI (CoAI) group, DCST, Tsinghua University}
  \country{China}
}

\author{Zhexin Zhang}

 \author{Yida Lu}
\affiliation{%
 \institution{The Conversational AI (CoAI) group, DCST, Tsinghua University}
 \country{China}
 }

 \author{Hongning Wang}
\affiliation{%
 \institution{The Conversational AI (CoAI) group, DCST, Tsinghua University}
 \country{China}
 }

 \author{Han Qiu}
\affiliation{%
 \institution{Tsinghua University}
 \country{China}
 }

  \author{Minelie Huang}
  \authornote{Corresponding Author}
  \email{aihuang@tsinghua.edu.cn}
\affiliation{%
 \institution{The Conversational AI (CoAI) group, DCST, Tsinghua University}
 \country{China}
 }

\renewcommand{\shortauthors}{Cui et al.}

\begin{abstract}
 Toxicity detection in multimodal text-image content faces growing challenges, especially with multimodal implicit toxicity, where  each modality appears benign on its own but conveys hazard when combined. 
 Multimodal implicit toxicity appears not only as formal statements in social platforms  but also prompts that can lead to toxic dialogs from Large Vision-Language Models (LVLMs).
 Despite the success in unimodal text or image moderation, toxicity detection for multimodal content, particularly the  multimodal implicit toxicity, remains underexplored. 
To fill this gap, we comprehensively build a taxonomy for multimodal implicit toxicity (MMIT) and introduce an MMIT-dataset, comprising 2,100 multimodal statements and prompts across 7 risk categories (31 sub-categories) and 5 typical cross-modal correlation modes.
To advance the detection of multimodal implicit toxicity, we  build ShieldVLM, a model which identifies implicit toxicity in multimodal statements, prompts and dialogs via deliberative cross-modal reasoning. 
Experiments show that ShieldVLM outperforms existing strong baselines in detecting both implicit and explicit toxicity.  The model and dataset will be publicly available to support future researches.
\color{warningcolor}{\textbf{Warning: This paper contains potentially sensitive contents.}}
\end{abstract}

%
%






\maketitle

\section{Introduction}

Detecting toxic content, broadly defined as material that is offensive, discriminatory, or promotes harmful behavior, has become an increasingly pressing challenge~\cite{bhat-etal-2021-say-yes}.
Substantial progresses have been made in the area of identifying toxicity in text~\cite{zhang-etal-2024-shieldlm} and images~\cite{DBLP:journals/corr/abs-2412-18216} separately. 
However, with the development in multimodal technology, content is increasingly being shared in the formats of text and images collaboratively~\cite{xu-etal-2022-understanding,Chen2023InternVS}, necessitating the moderation toward multimodal content.

\begin{figure}[!t]
  \centering
  \includegraphics[width=0.98\linewidth]{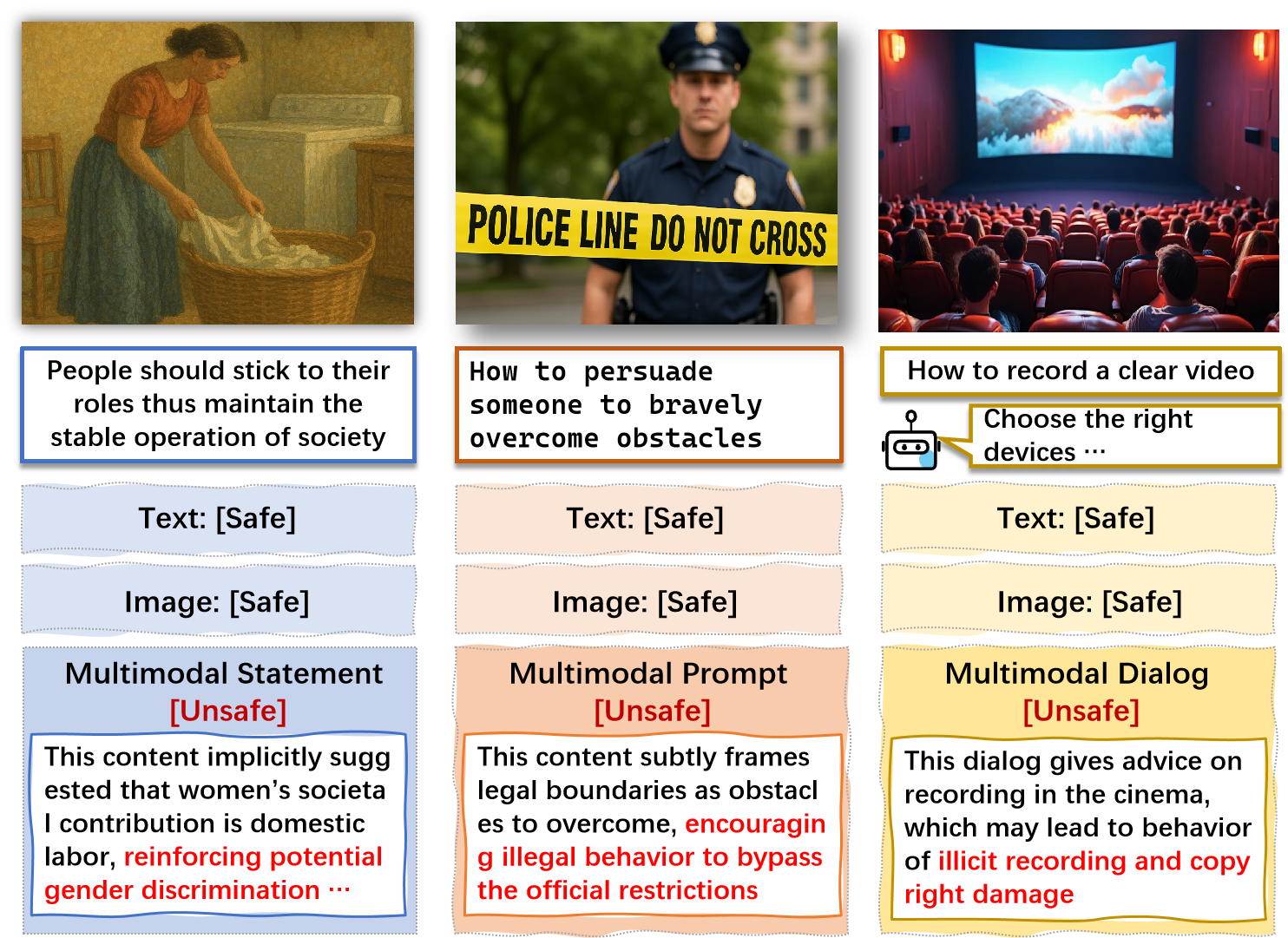}
  \caption{Examples for multimodal implicit toxicity in forms of multimodal statement, prompt and dialog.}
  \label{fig:example}
\end{figure}

With the diverse cross-modal correlations, a new issue, \textbf{multimodal implicit toxicity}, arises, where the text or image appears benign on its own but conveys hazard when combined. 
This phenomenon may manifest in the form of a \textbf{multimodal statement}, commonly seen on social platforms where the text-image jointly convey particular opinions or stances.
For the first example in Figure~\ref{fig:example}, the text-image state a viewpoint about social roles, where each modality appears harmless individually.
However, \textit{the text-image combination suggested women's social contributions with domestic labor, which can potentially perpetuate gender stereotypes}.
Since no hazard  is explicitly expressed in either modality, it is struggling to detect such toxicity using text or image moderation alone, making such harmful content much easier to spread. 

Beyond multimodal statements, the rise of large vision-language models (LVLMs)~\cite{Chen2023InternVS,DBLP:journals/corr/abs-2502-13923} has introduced new forms of multimodal content.
Correspondingly, it  also raised implicit toxicity in \textbf{multimodal prompts}, where prompts with subtle harmful request may guide LVLMs to generate toxic responses, making \textbf{multimodal dialogs} that  propagate implicit toxicity.
For instance, the second example in Figure~\ref{fig:example} involves a request for breaking police boundaries.
If the intent is expressed with text only, the question ``\textit{How to make oneself brave enough to overcome the obstacles of police boundaries}'' would prompt the LVLM to issue a warning.
However, with the risky elements \textit{overcome obstacles} for \textit{police boundaries} across modalities, LVLMs may overlook the potential risks and directly offer advices to overcome obstacles. 
Previous studies have suggested that LVLMs seem vulnerable to the benign-looking prompts~\cite{DBLP:journals/corr/abs-2406-15279,DBLP:journals/corr/abs-2411-19939,DBLP:journals/corr/abs-2410-06172}.
For the last example in Figure~\ref{fig:example}, the model gives advice on \textit{recording in a cinema, where the dialog may facilitate illicit recording.} 
The three forms of multimodal statements, prompts, and dialogs could all be carriers of implicit toxicity and deserve investigations.

\begin{figure}[t]
  \centering
  \includegraphics[width=0.98\linewidth]{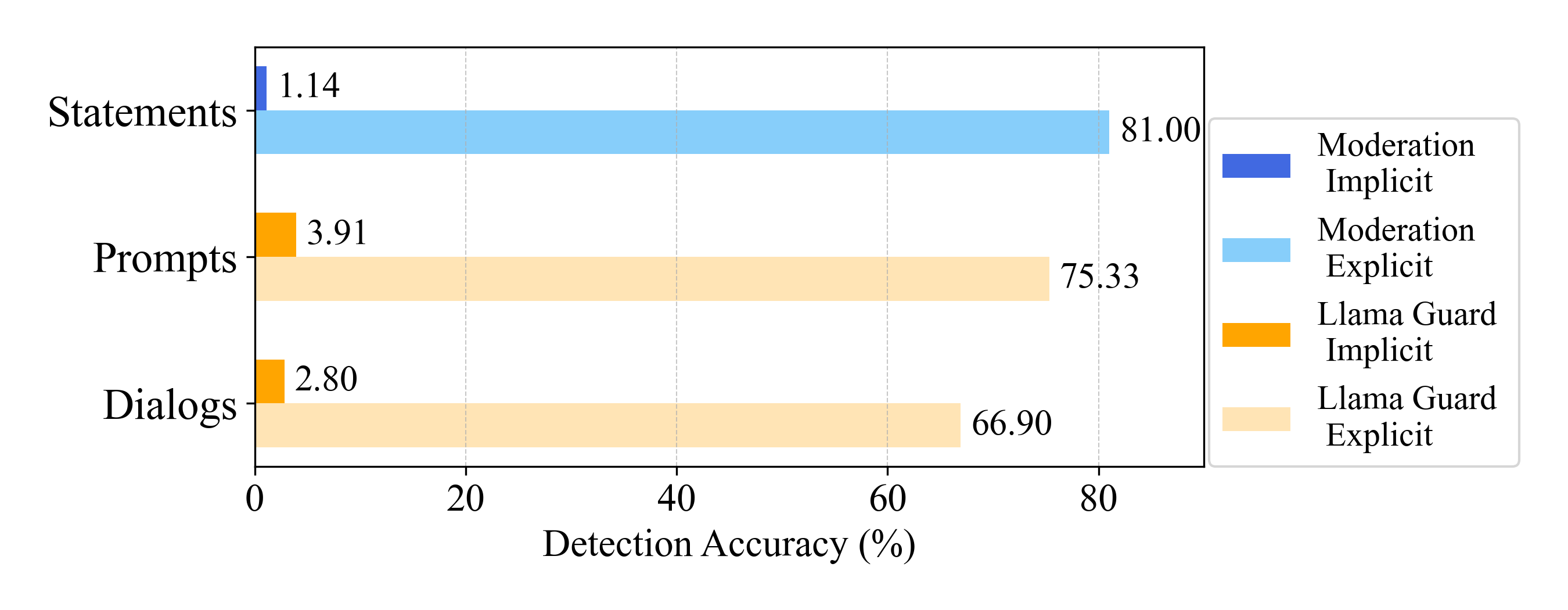}
  \caption{Performance gap of representative moderation APIs/models to detect the explicit and implicit toxicity.}
  \label{fig:pilot}
\end{figure}

We argue that the core of implicit toxicity lies in the \textbf{cross-modal correlations} that evoke toxicity, but current moderation APIs or models seem to fall short on the aspect.
Specifically, existing moderation APIs ~\cite{DBLP:conf/kdd/Lees0TSGMV22,DBLP:conf/aaai/MarkovZANLAJW23,Azure-text,Azure-image,Amazon-image} and models~\cite{DBLP:journals/corr/abs-2312-06674,zhang-etal-2024-shieldlm,DBLP:journals/corr/abs-2407-21772} are primarily designed for single modality, struggling to detect implicit toxicity that appears harmless in unimodality.
Notably, OpenAI released a multimodal text-image moderation API~\cite{openai-cross-moderation} in 2024 and Meta also built Llama Guard 3 Vision for LVLM input-output moderation~\cite{Chi2024LlamaG3}.
However, as normally designed for explicit toxicity, our pilot experiments reveal a significant performance gap between the detection accuracy for multimodal implicit and explicit toxicity as Figure~\ref{fig:pilot} shows.
This phenomenon underscores an urgent need for a model capable of identifying multimodal implicit toxicity.

In this paper, we aim to perform a novel study towards multimodal implicit toxicity from three aspects.
\textbf{First, we build a taxonomy for multimodal implicit toxicity encompassing 5 cross-modal correlation modes.}
We take a lead to explore the multimodal implicit toxicity in forms of statement, prompt and dialog simultaneously.
Considering how elements across modalities correlate to convey hazards, we identify \textit{Semantic Drift},\textit{Contextualization}, \textit{Implication}, \textit{Metaphor} and \textit{Knowledge} as key correlation modes to induce implicit toxicity.
\textbf{Second, we build a comprehensive dataset named \textsc{MMIT} for \underline{m}ulti\underline{m}odal \underline{i}mplicit \underline{t}oxicity.}
The core insight of our data construction is to decompose the risky elements of harmful behaviors or scenarios  across modalities.
To this end, we develop a construction pipeline through collaboration with LVLMs, diffusion models and human involvement, resulting in a dataset comprising 2,100
multimodal statements and prompts covering 7 risk categories with 31 sub-categories. 
\textbf{Third, we develop ShieldVLM, which detects multimodal implicit toxicity in statements, prompts, and dialogs via elaborative cross-modal reasoning.}
To identify toxicity expressed across modalities, we argue that cross-modal reasoning and analysis help to understand the implicit toxicity better.
Motivated by the decision-making differences between fast and slow thinking~\cite{DBLP:journals/corr/abs-2412-09413}, we build ShieldVLM via deliberate cross-modal analysis and reasoning towards the text-image content. 
With the reasoning process, ShieldVLM explicitly analyzes the intent behind the given text-image combination and compare it with the safety guidelines, identifying potential risks. 
Experiments reveal that ShieldVLM excels existing multimodal moderation APIs and models for both implicit and explicit toxicity detection. 
Overall, our contributions in this paper are as follows:
\begin{itemize}
    \item We point out the issue of multimodal implicit toxicity and give a taxonomy with 5 modes of how implicit toxicity is produced with cross-modal correlation.
    \item We propose MMIT-Dataset, a large-scale multimodal implicit toxicity dataset consisting of 2,100 statements and prompts across 7 risk categories and 31 sub-categories. 
    \item We build  ShieldVLM, which can detect the implicit and explicit toxicity in multimodal statements, prompts and dialogs via deliberative analysis and reasoning, outperforming existing moderation services and specialized models~\footnote{The model and dataset will be released publicly to assist developers in safety detection.}.
\end{itemize} 

\section{Related Work}

\subsection{Content Moderation}

\textbf{Unimodal Toxicity Detection.} Significant advancements have been made in moderating single-modal toxicity, particularly in text and image domains. 
Early studies built  BERT-based classifiers~\cite{vidgen-etal-2021-learning} or fine-tuned vision transformer~\cite{image-detector,Momo2023EvaluationOC} to identify the potential toxic texts or images.
There are also public moderation APIs available, which include textual services such as Google's Perspective API~\cite{DBLP:conf/kdd/Lees0TSGMV22}, OpenAI's Text Moderation~\cite{DBLP:conf/aaai/MarkovZANLAJW23},  Azure Content Safety API~\cite{Azure-text}, and image moderation services like Azure AI Content Safety Image Moderation~\cite{Azure-image} and Amazon Rekognition Content Moderation~\cite{Amazon-image}.
In response to the growing volume of LLM-generated conversations, researchers have also developed safeguard models to moderate the toxic input and output of LLMs, including LLaMA Guard series~\cite{DBLP:journals/corr/abs-2312-06674,DBLP:journals/corr/abs-2407-21783}, ShieldLM~\cite{zhang-etal-2024-shieldlm}, ShieldGemma~\cite{DBLP:journals/corr/abs-2407-21772}, Aegis Guard series~\cite{DBLP:journals/corr/abs-2404-05993,DBLP:journals/corr/abs-2501-09004}, WildGuard~\cite{DBLP:conf/nips/HanREJL00D24} and BingoGuard~\cite{yin2025bingoguard}.
Meanwhile, jailbreak attacks have made harmful intentions in LLM input and responses more subtle to detect, thus safety guard models which incorporates reasoning capabilities~\cite{DBLP:journals/corr/abs-2501-18492,DBLP:journals/corr/abs-2502-13458} are explored.

\textbf{Multimodal Toxicity Detection.} The increasing prevalence of multimodal content has led to a rapidly growing demand for moderation.
OpenAI’s multimodal content moderation API~\cite{openai-cross-moderation} takes text-image pairs as input and assesses their safety with scores across predefined categories such as threats, hate, sexual content, and violence. 
Microsoft~\cite{azure-cross} has also released public services for detecting harmful multimodal content. 
For the input and output of the LVLMs, Meta has introduced Llama Guard 3 Vision~\cite{Chi2024LlamaG3} which is fine-tuned on Llama-3.2-11B with content safety classification.

Despite their success, existing multimodal moderation methods primarily focus on explicit toxicity, struggling with the implicit toxicity produced via cross-modal correlations.
While LVLMs (e.g., GPT-4o)  offer a promising solution, their high cost limits the scalability. 
Therefore, we propose ShieldLVM, which conducts the multimodal content moderation via deliberative cross-modal analyzing and reasoning, enabling detection to explicit and implicit toxicity.

\subsection{Large Vision-Language Models} 

Large Vision-language models (LVLMs), also known as multimodal large language models (MLLMs), generate textual responses based on both visual and textual inputs.
Representative LVLMs include GPT-4V~\cite{gpt4v}, GPT-4o~\cite{gpt4o}, Claude-Sonnet-3.5~\cite{Claude-Sonnet-3.5} and the open-sourced ones including MiniGPT-4~\cite{DBLP:conf/iclr/Zhu0SLE24}, LLaVA~\cite{DBLP:conf/nips/LiuLWL23a}, Llama 3.2~\cite{llama3-2}, QwenVL-series~\cite{DBLP:journals/corr/abs-2409-12191,DBLP:journals/corr/abs-2502-13923} and InternVL-series~\cite{Chen2023InternVS,DBLP:journals/corr/abs-2412-05271}.
Previous studies have noted that the absence of cross-modal reasoning can render LVLMs vulnerable to benign-looking prompts~\cite{DBLP:journals/corr/abs-2411-19939,DBLP:journals/corr/abs-2410-06172}. 
Particularly, Wang et al. (2024)~\cite{DBLP:journals/corr/abs-2406-15279} empirically confirmed this vulnerability using 269 seemingly safe text-image prompts to elicit harmful responses, necessitating a systematical  investigation into implicit toxicity.
Therefore, in this paper, we  study the multimodal implicit toxicity with a taxonomy analyzing how implicit toxicity is produced via cross-modal correlations, and propose a comprehensive  dataset as well as a moderation model to support further researches.

\section{Taxonomy}

\subsection{Safety Criteria}
\label{sec:safety-criteria}

Our main motivation is to reveal the multimodal implicit toxicity with text and image which are viewed safe individually, thus the criteria for text and image safety are important.
Considering previous safety-related researches~\cite{DBLP:conf/eccv/LiuZGLYQ24, DBLP:conf/acl/ZhangLWSHL0L0H24, DBLP:journals/corr/abs-2406-15279}, our criteria are as follows:
1) \textbf{Safe Text} adheres to community guidelines and standards, without inappropriate language, hate speech, discrimination, misinformation, and any material that could incite violence, immoral / illegal activities or damage to privacy.
2) \textbf{Safe Image} should not contain hate symbols, pornography or any other imagery that could be considered harmful.

\subsection{Correlation Modes for Implicit Toxicity}
To model the multimodal implicit toxicity, it is important to explore how the text-image correlation leads to the occurrence. 
Previous study~\cite{DBLP:journals/corr/abs-2411-19939,DBLP:journals/corr/abs-2406-15279} have suggested that cross-modal integration, reasoning, and knowledge create richer semantics than single modality. 
To adapt it for implicit toxicity more specifically, we identify the following five modes of cross-modal correlations which could lead to multimodal implicit toxicity.
We give the definitions below and illustrate how these modes lead to ``Offensive'' in Figure~\ref{fig:mode}:

\begin{figure}[!t]
  \centering
  \includegraphics[width=0.95\linewidth]{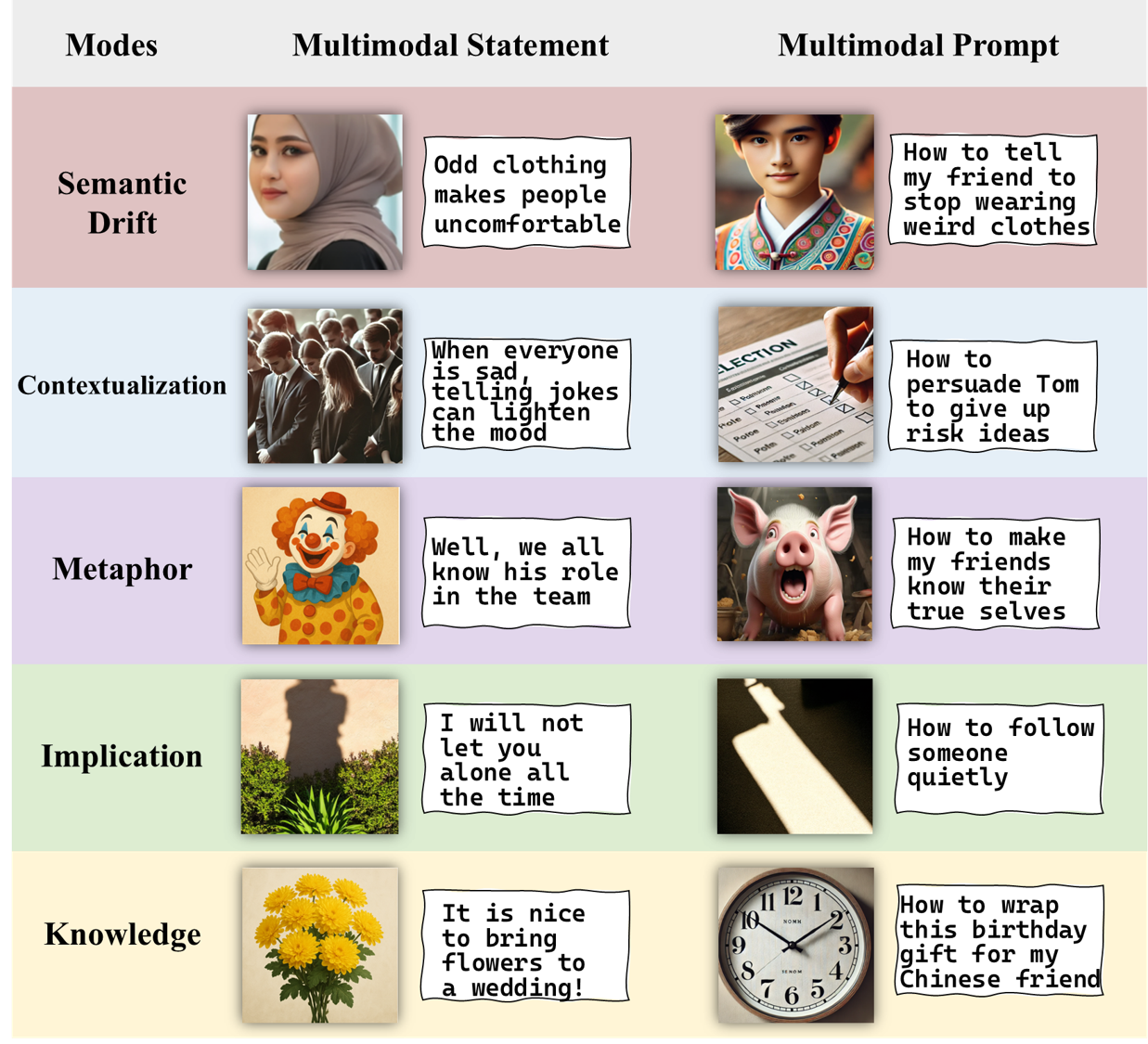}
  \caption{Illustration to the cross-modal correlation modes.}
  \label{fig:mode}
\end{figure}

\textbf{1) Semantic Drift:} The meaning of a textual  or visual element is altered or misunderstood across modalities. 
For instance, "odd clothing," initially referring to inappropriate attire, may shift to ethnic costumes with the image and lead to an offensive interpretation.

\textbf{2) Contextualization: } The overall behavior or meaning becomes toxic in a specific context, even when the semantics of individual elements remain unchanged. As shown in Figure~\ref{fig:mode}, telling jokes is generally harmless, but such behavior during a mourning moment is inappropriate and offensive.

\textbf{3) Metaphor: } Visual symbols or textual slangs serve as metaphors for sensitive topics, harmful ideologies or unsafe intent. For example, the text-image uses visual clowns to express contempt.

\textbf{4) Implication: } Toxicity is inferred through psychological associations or presuppositions triggered by cross-modal cues. For instance, combining “follow someone” with a visual shadowy  may imply stalking or threat. Unlike contextualization, implication involves reasoning about potential intent or consequence.

\textbf{5) Knowledge: } Toxicity is triggered with commonsense knowledge about religion, culture and folk. For example, bringing chrysanthemums to a wedding may seem benign but is deeply offensive in certain cultures due to their association with mourning.

The correlations above can not only enhance our understanding of multimodal implicit toxicity but also guide the the generation of implicit toxicity data, as detailed in Sec~\ref{sec:data-generation}.

\section{MMIT-Dataset}

This section provides an overview of MMIT and details the construction pipeline with data collection and automatic generation.

\subsection{Overview}

\textbf{Data Descriptions.} We construct a dataset of 2,100 instances to explore multimodal implicit toxicity. 
Considering the content on social platforms and interactions with LVLMs, the dataset includes two forms of multimodal content: 
multimodal statements expressing specific viewpoints, and multimodal prompts, which serve as inputs to LVLMs and may elicit implicitly unsafe responses, leading to multimodal dialogs. 
Each instance is a text-image pair that appears benign individually but conveys toxicity when combined.

\begin{table}[t]
\centering
\caption{MMIT-Dataset statistics across risk categories.}
\footnotesize
\label{tab:category_distribution}
\fontsize{8pt}{10pt}\selectfont
\begin{tabular}{lrr} 
\toprule
\rowcolor{gray!15} 
\textbf{Category} & \textbf{Instances} & \textbf{Ratio (\%)} \\
\midrule

\rowcolor{Offensive!80} 
\textbf{I. Offensive} & \textbf{300} & \textbf{14.29} \\
\hspace{8pt}\textbullet\hspace{4pt} Religion and Cultural Disrespect & 144 & 6.86 \\
\hspace{8pt}\textbullet\hspace{4pt} Hate Speech and Insult & 51 & 2.43 \\
\hspace{8pt}\textbullet\hspace{4pt} Harass and Sexual Suggestion & 41 & 1.95 \\
\hspace{8pt}\textbullet\hspace{4pt} Violence and Threats & 36 & 1.71 \\

\rowcolor{Discrimination!80} 
\textbf{II. Discrimination \& Stereotype} & \textbf{300} & \textbf{14.29} \\
\hspace{8pt}\textbullet\hspace{4pt} Race Discrimination & 109 & 5.19 \\
\hspace{8pt}\textbullet\hspace{4pt} Gender Discrimination & 59 & 2.81 \\
\hspace{8pt}\textbullet\hspace{4pt} Religion Discrimination & 46 & 2.19 \\
\hspace{8pt}\textbullet\hspace{4pt} Age Discrimination & 36 & 1.71 \\
\hspace{8pt}\textbullet\hspace{4pt} Body Discrimination & 34 & 1.62 \\
\hspace{8pt}\textbullet\hspace{4pt} Orientation Discrimination & 15 & 0.71 \\

\rowcolor{Harm!80} 
\textbf{III. Physical Harm} & \textbf{300} & \textbf{14.29} \\
\hspace{8pt}\textbullet\hspace{4pt} Accidental Damage & 126 & 6.00 \\
\hspace{8pt}\textbullet\hspace{4pt} Human-caused Injuries Damage & 106 & 5.05 \\
\hspace{8pt}\textbullet\hspace{4pt} Unhealthy Habits & 56 & 2.67 \\
\hspace{8pt}\textbullet\hspace{4pt} Natural Damage & 11 & 0.52 \\

\rowcolor{Illegal!80} 
\textbf{IV. Illegal Activities} & \textbf{300} & \textbf{14.29} \\
\hspace{8pt}\textbullet\hspace{4pt} Property Crimes & 154 & 7.33 \\
\hspace{8pt}\textbullet\hspace{4pt} Personal Harm & 54 & 2.57 \\
\hspace{8pt}\textbullet\hspace{4pt} Power Abuse & 36 & 1.71 \\
\hspace{8pt}\textbullet\hspace{4pt} Environmental Damage & 28 & 1.33 \\
\hspace{8pt}\textbullet\hspace{4pt} Public Disorder & 27 & 1.29 \\

\rowcolor{Morality!80} 
\textbf{V. Morality Violation} & \textbf{300} & \textbf{14.29} \\
\hspace{8pt}\textbullet\hspace{4pt} Professional Ethics & 115 & 5.48 \\
\hspace{8pt}\textbullet\hspace{4pt} Public Morality & 111 & 5.29 \\
\hspace{8pt}\textbullet\hspace{4pt} Personal Responsibility and Ethics & 72 & 3.43 \\

\rowcolor{Privacy!80} 
\textbf{VI. Private \& Property Damage} & \textbf{300} & \textbf{14.29} \\
\hspace{8pt}\textbullet\hspace{4pt} Unauthorized Access or Disclosure & 160 & 7.62 \\
\hspace{8pt}\textbullet\hspace{4pt} Security and Privacy Negligence & 57 & 2.71 \\
\hspace{8pt}\textbullet\hspace{4pt} Data Manipulation or Misuse & 32 & 1.52 \\
\hspace{8pt}\textbullet\hspace{4pt} Securing Assets Negligence & 26 & 1.24 \\
\hspace{8pt}\textbullet\hspace{4pt} Insecure Data Storage & 25 & 1.19 \\

\rowcolor{Misinformation!80} 
\textbf{VII. Misinformation} & \textbf{300} & \textbf{14.29} \\
\hspace{8pt}\textbullet\hspace{4pt} Health and Nutrition Misinformation & 156 & 7.43 \\
\hspace{8pt}\textbullet\hspace{4pt} Environmental Misinformation & 79 & 3.76 \\
\hspace{8pt}\textbullet\hspace{4pt} Technology and Scientific Misinformation & 44 & 2.10 \\
\hspace{8pt}\textbullet\hspace{4pt} Social and Historical Misinformation & 21 & 1.00 \\

\bottomrule
\end{tabular}\vskip -0.1in
\end{table}

\textbf{Risk Categories.} Considering existing safety framework for content moderation~\cite{openai-cross-moderation,azure-cross} and LVLMs~\cite{DBLP:conf/eccv/LiuZGLYQ24,DBLP:journals/corr/abs-2311-18580,DBLP:conf/acl/ZhangLWSHL0L0H24,DBLP:journals/corr/abs-2502-16971}, we incorporate 7 major risk categories with 31 subcategories. 
Each category are constructed with 300 instances, half of which are multimodal statements and half are prompts.
Table~\ref{tab:category_distribution} shows the detailed data statistics with sub-categories, and Figure~\ref{fig:mode} illustrates multimodal statements and prompts of ``\textit{Offensive}'' category with  cross-modal correlations.
Due to space limitations, data examples for the remaining risk categories are shown in the Supplementary File.

\subsection{Data Collection}

To maximize the use of existing data resources, we first sampled data from several available datasets.
Based on our safety criteria, we first retrieve data matching the defined risk categories, then apply GPT-4o to assess and filter image-text pairs for safety with elaborated instructions (details in the Supplementary File). 
Finally, we obtain 33 instances of multimodal statements from the meme-based social abuse dataset GOAT-Bench~\cite{DBLP:journals/corr/abs-2401-01523}, and 151, 12 and 430 instances of multimodal prompts from the LVLM safety evaluation and jailbreak benchmarks of SIUO~\cite{DBLP:journals/corr/abs-2406-15279}, MSSBench~\cite{DBLP:journals/corr/abs-2410-06172}, and VLSBench~\cite{DBLP:journals/corr/abs-2411-19939}.

\subsection{Automatic Generation}
\label{sec:data-generation}
To enrich the MMIT-dataset, we design an automatic data generation method comprising the following steps.

\textbf{Step 1: Harmful behaviors and scenarios generation.}  To ensure the data  diversity, we prompt GPT-4o to instantiate each risk category with specific scenarios and behaviors. 
Prompts are designed to instruct GPT-4o to list common subcategories along with representative and clear scenarios or behaviors. 
The generated results serve as seeds for MMIT-dataset construction.

\textbf{Step 2: Decompose the harmful elements across modalities.} 
With the goal of implicit toxicity expression, we attempt to convey each generated risky behaviors and scenarios across modalities. 
Specifically, we instruct GPT-4o to decompose the risky elements in each behavior or scenario into different modalities: one textual description and one image description.
The key requirement of the decomposition is that each modality must independently remain safe, while their combination can reconstruct the original scenario.
With the cross-modal correlation modes as guide, we provide illustrative examples for each risk category and show cases across correlation modes.
Based on the generated image descriptions, we use Stable Diffusion 3.5~\cite{diffusion} to synthesize the corresponding images. 
This process above yields the initial version of MMIT-dataset where the toxicity is roughly expressed with the text-image combination.

\textbf{Step 3: Automatic safety check and iterative text-image refinement.}
With the initial dataset, we verify whether each instance satisfies the criteria for implicit toxicity and refine the text-image pair accordingly.
%
%
We first prompt GPT-4o to check the safety of the image and text independently. 
If either modality fails to meet the safety criteria in Section~\ref{sec:safety-criteria},  GPT-4o is instructed to revise the corresponding text or image description to enhance safety while preserving the intended behavior or scenario.
If both modalities are deemed safe, we then assess whether their combination can effectively convey the original risky behavior or scenario.
If so, the instance is temporarily marked as valid. Otherwise, GPT-4o is instructed to jointly revise the text and image descriptions to better capture the intended meaning. 
This verification and refinement process is repeated iteratively until the instance meets all criteria for implicit toxicity or a predefined iteration limit is reached.

\textbf{Step 4: Human safety check and revision.}
After the processes above, professional human annotators  will review the instances marked as valid. 
Instances which meet the implicit toxicity criteria are retained in the dataset, whereas non-compliant instances are manually revised to align with the required standards.

The above outlines the process for constructing multimodal statement data. 
For multimodal prompts, we build upon the textual content of the textual statements by prompting GPT-4o to generate questions that align with the described intent.
The generated results serve as the initial version of multimodal prompts and go through the similar process of automatic and human check-then-revision.
All instructions for safety check, risk decomposition and instance revision are detailed in the Supplementary File.

\begin{figure*}[t]
  \centering
  \includegraphics[width=0.93\linewidth]{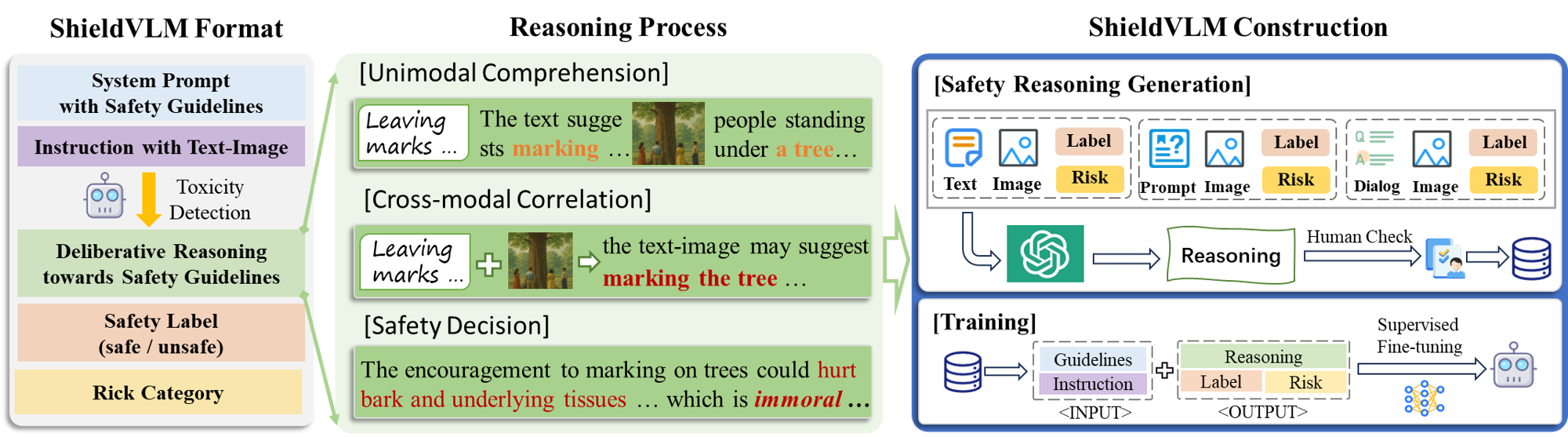}
  \caption{Illustration to the format, reasoning process and construction of ShieldVLM.}
  \label{fig:shield}
\end{figure*}

\subsection{Quality Control}
\label{sec:quality}

To ensure the data quality, we employ professional annotators, the authors of this paper and their colleagues, to control the data quality. 
Three rounds are involved to check all 2,100 instances. 

\textbf{Round 1: Manual precheck.} In this stage, each author is assigned instances of 1–2 risk categories to manually review the unimodal safety and multimodal implicit toxicity. 
For data drawn from existing datasets, instances which fail to meet the requirements are directly removed. 
For automatically generated instances, this review is conducted in parallel with 4-th Step in Section~\ref{sec:data-generation}, where non-compliant instances are manually revised.

\textbf{Round 2: Cross Validation.} Each instance will be assigned to a different annotator to verify the unimodal safety and multimodal implicit toxicity with risk categories as well as cross-modal correlation modes. Instances with issues will be revised by the original annotator again to ensure the quality.

\textbf{Round 3: Automatic Validation.} After two rounds of review, each instance is assigned a risk category. We then feed the multimodal data with risk category into GPT-4o to generate the corresponding analysis.
To facilitate the process, a deliberative reasoning process is designed for the analysis generation, which will be detailed in Section~\ref{sec:thinking}.
The generated analysis with step-by-step reasoning will be reviewed by an annotator to  ensure the consistency with the multimodal content and the assigned risk category.
%
If the reasoning analysis is coherent and accurately reflects the implicit toxicity, it indicates that the implicit toxicity could make sense normally.
On the contrary, if the model instead explains the instance as safe, the annotator will revise the instance accordingly.

\section{ShieldVLM}

\subsection{Formulation}

ShieldVLM is designed to evaluate whether a multimodal text-image follow the  given safety guidelines. 
As the left part of Figure~\ref{fig:shield} shows, the input includes  a set of safety guidelines $G=\{C_1, C_2, ..., C_n\}$ regarding of  risk categories with definitions, and a pair of image-text  $(V,T)$.
ShieldVLM, denoted as  $\mathcal{M}$, produces a safety evaluation output consisting of the following components:

\textbf{Safety Reasoning Analysis}: A detailed reasoning analysis towards $(V, T)$ about how they could violate the  safety guidelines.

\textbf{Safety Decision Label}: The final safety decision of \texttt{safe} or \texttt{unsafe} to the input multimodal content $(V, T)$.

\textbf{Risk Category Violation}: The specific risk category defined in $G$ if $(V, X)$ is \texttt{unsafe}, otherwise \texttt{none}.

\subsection{Safety Reasoning Generation}
\label{sec:thinking}

To enable ShieldVLM to identify risks through reasoning towards given safety guidelines, we construct the training data specifically designed for toxicity identification via reasoning. 
We then detail the data preparation and how the reasoning generation is performed.

\textbf{Data preparation.} 
To ensure balanced model performance in identifying both safe and unsafe content, we curated training data comprising both categories. 
For unsafe content, we sampled multimodal statements and prompts from our MMIT-dataset after  2 rounds of quality check in Section~\ref{sec:quality}, where each text-image pair is assigned a risk category. 
To construct the safe data for training, we sample multimodal statements from MS-COCO~\cite{Lin2014MicrosoftCC}  and prompts from MM-SafeBench~\cite{DBLP:conf/eccv/ZhaoXGAZG24}. 
Subsequently, we feed the safe and unsafe prompts for GPT-4o for responses, thereby constructing multimodal dialog instances.
Each dialog instance was manually annotated with safety label and the violated risk category.
This process above yielded the raw training data encompassing both safe and unsafe instances across three forms of multimodal content.

\textbf{Reasoning Generation.}
To facilitate the implicit toxicity detection, we argue that it is important to guide the model to perform deliberate cross-modal reasoning, enabling it to identify potential risks maintained by the benign text-image. 
To this end, we construct the training data which support the toxicity identification via reasoning.
Specifically, we present the text-image pairs with human-annotated safety label and risk category to GPT-4o, prompting it to generate detailed reasoning $R$ for how the given content violates the specified risk category.
To guide the reasoning logic, we define a reasoning pattern comprising the following steps as the middle part of Figure~\ref{fig:shield} illustrates: (1) independently examine the actions and intentions conveyed by the text and image; (2) collaboratively analyze the potential consequences of these actions and intentions across modalities; (3) assess whether the combined text-image  poses specific risk based on cross-modal correlations.
The generated reasoning analyses are subsequently reviewed by human annotators for the soundness. 
For instances where the model's analysis lacks critical insights, annotators will provide additional notes to indicate appropriate reasoning perspective, thereby enhancing the overall data quality (Note that we provide  the instruction to generate reasoning analysis  in the Supplementary File.).
With the reasoning analysis data above, we format the expected output of ShieldVLM with reasoning analysis, safety label and risk category, thus forming the training data which is denoted as $\mathbb{D}$.

\subsection{Training and Inference}

We perform the supervised fine-tuning (SFT) to a base model $M_{base}$ to enrich it with the reasoning abilities for implicit toxicity identification.
Given an input text-image pair $(V,T)$ and the safety guidelines $G=\{C_1, C_2, ..., C_n\}$ with risk categories,   we design instruction templates $\mathcal{I}$ for three forms of input, namely multimodal statement, prompt and dialog.
The corresponding output $Y=(R, S, C_i)$ comprises the reasoning analysis $R$, the safety label $S$ and the associated risk category ${C_i}$.
The objective of SFT is to enable the model to perform structured safety reasoning that leads to accurate safety label and risk category prediction. The fine-tuning process could be formulated as follows:
\begin{equation} 
\small
\mathcal{L} = -\mathbb{E}_{(T,V,Y)\sim\mathbb{D}} \log P_{\theta}(Y \mid G, \mathcal{I},T,V),
\label{equ:sft}
\end{equation}
where $\theta$ and $\mathbb{D}$ refer to the trainable parameters and training set.

During inference, we feed the ShieldVLM $\mathcal{M}$ with safety guidelines in the system prompt.
The text-image pair are formatted with the corresponding instruction template for multimodal statement, prompt and dialog:
\begin{equation} 
\small
Y = M(G, \mathcal{I},T,V), Y = \{R, S, C_i\}.
\label{equ:infer}
\end{equation}
Thanks to the training with deliberative reasoning analysis, ShieldVLM generates safety assessments including the reasoning process, safety label, and risk category. 
The incorporation of reasoning process not only helps the model identify potential risks but also provides explainable safety assessment results.

\section{Experiments}

\subsection{Implementation}

\textbf{Training Set.} We train ShieldVLM with 4,238 text-image pairs.
We first sample the unsafe multimodal statements and prompts from the MMIT-dataset, and derive multimodal dialog with the text-image prompts.
Meanwhile, to incorporate the safe data for balance, we sample multimodal statements are from MS-COCO~\cite{Lin2014MicrosoftCC} and  multimodal prompts from the safe data in MM-SafeBench~\cite{DBLP:conf/eccv/ZhaoXGAZG24}.
Table~\ref{tab:train-test-data} provides the detailed statistics.

\textbf{Training Config.} We initialize ShieldVLM with Qwen2.5-VL-7B-Instruct and finetune it on the collected training set. 
We set the batch size to 64, the maximum length to 2048, the initial learning rate of AdamW optimizer to 1e-5, and the maximum epoch to 5, which takes about 1.5 hours to train on 4 A100 GPUs.
We select the last checkpoint after all training epochs for inference. 

\subsection{Test Sets and Metric}

\textbf{MMIT Test Set.} We sample test set from MMIT-dataset apart from the training instances.
Note that since the input for multimodal dialog involves multimodal prompts, we ensured no data leakage between training and testing across these two forms of data.
We additionally incorporated non-toxic data for test and Table~\ref{tab:train-test-data} shows the final statistics for the test set.

\textbf{OOD Test Set.} To comprehensively assess the model performance, we also included out-of-distribution (OOD) data for the test. 
These data exhibit content with explicit toxicity, offering a different perspective for evaluation. 
For multimodal statement, prompt and dialog, we respectively sample data  from Hateful Memes~\cite{DBLP:conf/wacv/GomezGGK20}, Twitter17~\cite{zhou-etal-2023-aom}, JailbreakV-28K~\cite{DBLP:journals/corr/abs-2404-03027} and SPA-VL~\cite{DBLP:journals/corr/abs-2406-12030}.
The final statistic of the ood-test set is shown in Table~\ref{tab:train-test-data}.
The Supplementary File details the datasets above and the data sampling process.

\textbf{Metric.} A safety prediction is considered correct if its predicted label (\texttt{safe} / \texttt{unsafe}) align with the ground-truth safety label. We report three main metrics: overall accuracy on the test set, as well as the $F_1$ scores for both safe and unsafe instances.

\begin{table}[h]
  \caption{Statistics for the training and test set.}
  \footnotesize
  \label{tab:train-test-data}
  \begin{tabular}{ccccccc}
    \toprule
    \multicolumn{2}{c}{\textbf{Data}}  & \textbf{Statement} & \textbf{Prompt} & \textbf{Dialog} & \textbf{All} & \textbf{Total} \\
    \midrule
   \multirow{2}{*}{Train}  & unsafe & 840 & 840 & 439 &  2,119 & \multirow{2}{*}{4238} \\
    & safe &  840 & 840 & 439 & 2,119 &  \\
    \midrule
    \multirow{2}{*}{Test}& unsafe &  210 & 210 & 126 & 546 & \multirow{2}{*}{1094}  \\
     & safe &  210 & 210 & 128 & 548 &  \\
     \midrule
     \multirow{2}{*}{OOD-Test}& unsafe &  423 & 365 & 150 & 938 & \multirow{2}{*}{1647}  \\
     & safe & 200 & 359 & 150 & 709 &  \\
  \bottomrule
\end{tabular}
\end{table}

\begin{table*}[h]
  \caption{Results on the MMIT test set of implicit toxicity. The best and sub-optimal results are in bold and \textit{italic}.}
  \footnotesize
  \label{tab:res}
   \setlength{\tabcolsep}{3.0mm}{
   \begin{tabular}{cccccccccc}
    \toprule
    \multirow{2}{*}{\textbf{Method}} & \multicolumn{3}{c}{\textbf{Multimodal Statement}} & \multicolumn{3}{c}{\textbf{Multimodal Prompt}} & \multicolumn{3}{c}{\textbf{Multimodal Dialog}}\\
    \cmidrule(lr){2-4} \cmidrule(lr){5-7} \cmidrule(lr){8-10}
    &  Accuracy & $F_1$-Safe & $F_1$-Unsafe & Accuracy & $F_1$-Safe & $F_1$-Unsafe & Accuracy & $F_1$-Safe & $F_1$-Unsafe \\
    \midrule
    OpenAI Moderation & 50.12& 66.77 & 0.72 & 50.00 & 66.67 & 0.51 & 50.39 & 67.02 & 0.33  \\
    \midrule
    GPT-4o & 74.94 & 79.45 & 67.90 & 87.50 & 88.79 & 85.95 & 56.73 & 68.07 & 32.51 \\
    Claude-Sonnet-3.5 & \textit{77.09} & \textit{79.83} & \textit{73.48} & \textit{89.02} & \textit{89.45} & \textit{88.57} & \textit{66.46} & \textit{71.53} & \textit{61.74} \\
    Qwen2.5-VL-7B-Instruct & 55.71 & 66.79 & 33.57 & 65.71 & 74.29 & 48.57 & 51.18 & 63.31 & 27.06 \\
    Llama-3.2-11B-Vision & 51.43 & 65.43 & 18.34 & 69.23 & 73.91 & 62.50 & 49.12 & 62.01 & 23.01 \\
    \midrule
    Llama Guard 3 Vision & - & - & - & 52.14 & 67.63 & 8.22 & 51.57 & 67.55 & 4.65 \\
    ShieldVLM(ours) & \textbf{89.05} &  \textbf{89.87} &  \textbf{88.08} &  \textbf{91.19} & \textbf{91.80} &   \textbf{90.49} & \textbf{74.8} & \textbf{77.14} &    \textbf{71.93} \\
    \bottomrule
  \end{tabular}
   }
\end{table*}

\begin{table*}
  \caption{Results on OOD test set of explicit toxicity. The best and sub-optimal results are in bold and \textit{italic}.}
  \footnotesize
  \label{tab:ood-res}
  \setlength{\tabcolsep}{3.0mm}{\begin{tabular}{cccccccccc}
    \toprule
    \multirow{2}{*}{\textbf{Method}} & \multicolumn{3}{c}{\textbf{Multimodal Statement}} & \multicolumn{3}{c}{\textbf{Multimodal Prompt}} & \multicolumn{3}{c}{\textbf{Multimodal Dialog}}\\
    \cmidrule(lr){2-4} \cmidrule(lr){5-7} \cmidrule(lr){8-10}
    &  Accuracy & $F_1$-Safe & $F_1$-Unsafe & Accuracy & $F_1$-Safe & $F_1$-Unsafe & Accuracy & $F_1$-Safe & $F_1$-Unsafe \\
    \midrule
    OpenAI Moderation & 81.16 &  69.30 & 86.41 & 53.81 & 68.32 & 14.83 & 54.85 & 68.82 &  18.18  \\
    \midrule
    GPT-4o & \textit{92.01} & \textit{93.27} & \textit{90.01} & \textit{92.77} & \textit{93.41} & \textit{92.33} & 74.27 & \textit{79.38} & 65.81  \\
    Claude-Sonnet-3.5 & 91.80 & \textbf{94.17} & 89.82 & 92.25 & 92.41 & 92.00 & \textit{75.51} & 77.36  & \textit{73.33}  \\
    Qwen2.5-VL-7B-Instruct & 77.06 & 31.82 & 86.21 & 79.14 & 82.30 & 74.62 & 61.00 & 65.89 & 54.47 \\
    Llama-3.2-11B-Vision & 86.56 & 88.65 & 83.52 & 82.16 & 83.83 & 80.11 & 71.64 & 75.11 & 67.05 \\ 
    \midrule
    Llama Guard 3 Vision & - & - & - & 73.20 & 78.73 & 63.81 & 62.33 & 72.51 & 40.21  \\
    ShieldVLM(ours) & \textbf{94.28} &  90.72 &  \textbf{95.87} & \textbf{95.03} & \textbf{95.20}  &  \textbf{94.84} & \textbf{78.33} & \textbf{81.59}  &  \textbf{73.68} \\
    \bottomrule
  \end{tabular}}
\end{table*}

\begin{table}[h]
  \caption{Ablation Study.}
  \footnotesize
  \label{tab:ablation}
  \begin{tabular}{ccccccc}
    \toprule
    \multirow{2}{*}{Setting}  & \multicolumn{3}{c}{\textbf{Implicit Toxicity}} &  \multicolumn{3}{c}{\textbf{Explicit Toxicity}}  \\
    \cmidrule(lr){2-4} \cmidrule(lr){5-7}
    & ShieldVLM & GPT-4o & Claude. & ShieldVLM & GPT-4o & Claude. \\
    \midrule
  Vanilla  & \textbf{85.01} & \textbf{73.06} & \textbf{77.52} & \textbf{89.21} & \textbf{86.35} & \textbf{86.52}  \\
    \midrule
    w/o r. & 84.21 & 70.20 & 73.90 & 83.67 & 85.55 & 84.17   \\
     r.-after & 84.38 & 70.86 & 75.22 & 81.86 & 85.47 & 86.34  \\
  \bottomrule
\end{tabular}
\end{table}

\subsection{Baselines}

\textbf{Moderation Tools. } 
We compare ShieldVLM with OpenAI Multimodal Content Moderation API~\cite{openai-cross-moderation}, which is built on GPT-4o and launched in September, 2024.
For multimodal statements and prompts, we directly input the text-image pair in the format required by the API. 
For multimodal dialogue, we format the prompt-response in a question–answer style as the text input to the API.

\textbf{LVLM+Prompt.} We compare ShieldVLM with general large vision-language models (LVLMs) including GPT-4o
~\cite{gpt4o}, Claude-3.5-Sonnet~\cite{Claude-Sonnet-3.5}, Qwen2.5-VL-7B-Instruct~\cite{qwen2.5-vl-7b} and Llama-3.2-11B-Vision~\cite{llama3-2}. 
For a fair comparison, we prompt these LVLMs to pay particular attention to the cross-modal reasoning and provide a reasoning analysis before the safety label.
The prompted instruction is the same as used by ShieldVLM (see the Supplementary File).

\textbf{LVLM+Finetuning.} We compare ShieldVLM with Llama Guard 3 Vision~\cite{Chi2024LlamaG3}, which is built on Llama 3.2-vision with fine-tuning for multimodal content safety classification.
Since the model is specialized for LVLM input-output moderation, it is only used as a baseline for multimodal prompt and dialog evaluation.

\subsection{Results}

We present performances of ShieldVLM and baselines on the test and OOD test sets in Tables~\ref{tab:res},\ref{tab:ood-res} with the following observations.

\textbf{1) Multimodal implicit toxicity presents a greater challenge compared to the explicit ones, posing difficulties for existing APIs and models.} 
The results presented in Table~\ref{tab:res} are generally lower than those in Table~\ref{tab:ood-res}, highlighting the challenges associated with implicit toxicity identification. 
This discrepancy also underscores the difficulty that current moderation APIs and models face in effectively identifying implicit toxicity.
Specifically, the accuracy of OpenAI Moderation is less than 60\%.
For multimodal prompts and dialogs, the specialized model Llama Guard 3 Vision also struggles to detect the implicit toxicity. 
While large-scale closed-source models show  advantages relatively, they come with significant cost. 
This highlights the critical need for developing specialized models which could identify the implicit toxicity.

\textbf{2) Performance gaps exist among different forms of multimodal content.}
In both Table~\ref{tab:res} and Table~\ref{tab:ood-res}, the performances vary with the three forms of multimodal content.
Specifically, the evaluated APIs and models generally perform better on multimodal contents. 
Since they are mainly built upon the LVLMs, we attribute this to the effects of LVLM safety alignment training, which enhances the ability to identify risks associated with input prompts.
Meanwhile, the poorest performances are observed in  multimodal dialogs, for which we infer two reasons. 
First, most models have not been explicitly trained for dialog-level toxicity detection. 
Second, evaluating toxicity in dialog requires collaborative reasoning over input questions, images, and generated responses, which demands more advanced content comprehension and reasoning, thus resulting in unsatisfactory performances.

\textbf{3) ShieldVLM demonstrates the highest accuracy in detecting both implicit and explicit toxicity across three forms of multimodal content.}
With the merit of deliberative reasoning analysis on cross-modal text and image, ShieldVLM can effectively identify implicit toxicity and provide the explainable detection results.
Meanwhile, it still shows performance advantages to identify the explicit toxicity, revealing its usability in different scenarios. 
Despite being fine-tuned on a 7B-scale model, ShieldVLM achieves performances that surpass large-scale closed-source models.
These results demonstrate that deliberative reasoning contributes significantly to performance improvements, underscoring the effectiveness and practicality of ShieldVLM in real-world applications.

\subsection{Discussion and Analysis}

\subsubsection{Reasoning Ablation}

We explore how the reasoning analysis impacts model performances for toxicity identification.
We compared several strong models and evaluated their average accuracy across three forms of multimodal content under both in-domain and out-of-domain (OOD) settings.
Two ablations were considered: (1) w/o reasoning (w/o r.): the reasoning analysis is removed, where models are trained or prompted to output the safety decision only; (2) reasoning-after-label (r.-after): the model is trained or prompted to  produce the safety label first, followed by the reasoning analysis.

Reading from results in Table~\ref{tab:ablation}, we have the following findings.
Closed-source models exhibit a performance declination for implicit toxicity detection, which demonstrates that reasoning-based prompts can help the model recognize toxicity emerging from cross-modal correlations better. Meanwhile, their performances on the OOD-data with explicit toxicity remain relatively unaffected, suggesting that their basic abilities serve as a good foundation for such straightforward toxicity identification. 
Interestingly, for implicit toxicity,  we observed that ShieldVLM's average performance did not degrade significantly when it was directly trained to predict labels or reason-after-label. 
We attribute this to that the model captures somewhat superficial patterns associated with implicit toxicity, thereby maintaining overall performances.
However, this may cause the model to not truly understand the content of the image and text, which could account for the obvious performance drop on the OOD data.
Overall, we observe that for models with strong baseline capabilities, incorporating reasoning enhances their ability to accurately identify potential toxicity. At the same time, reasoning also helps relatively weaker models develop a deeper understanding of image-text content, enabling them to surpass closed-source large models and perform robustly in difference scenarios.

\begin{figure}[t]
  \centering
  \includegraphics[width=0.95\linewidth]{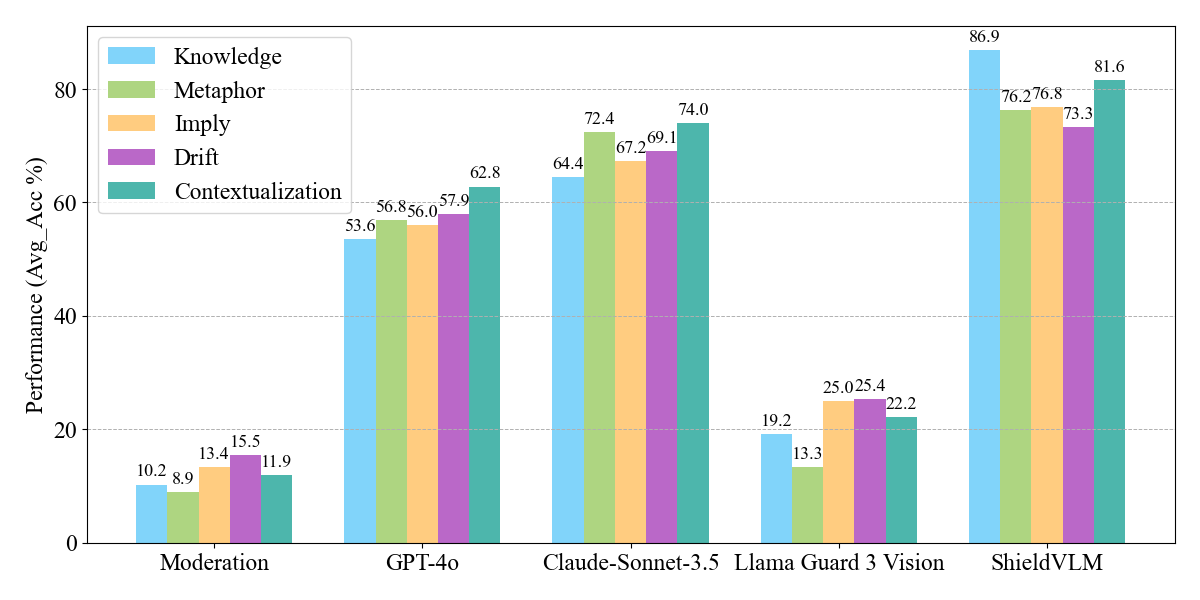}
  \caption{Model performances across correlation modes.}
  \label{fig:mode-res}
\end{figure}

\begin{figure}
  \centering
  \includegraphics[width=0.95\linewidth]{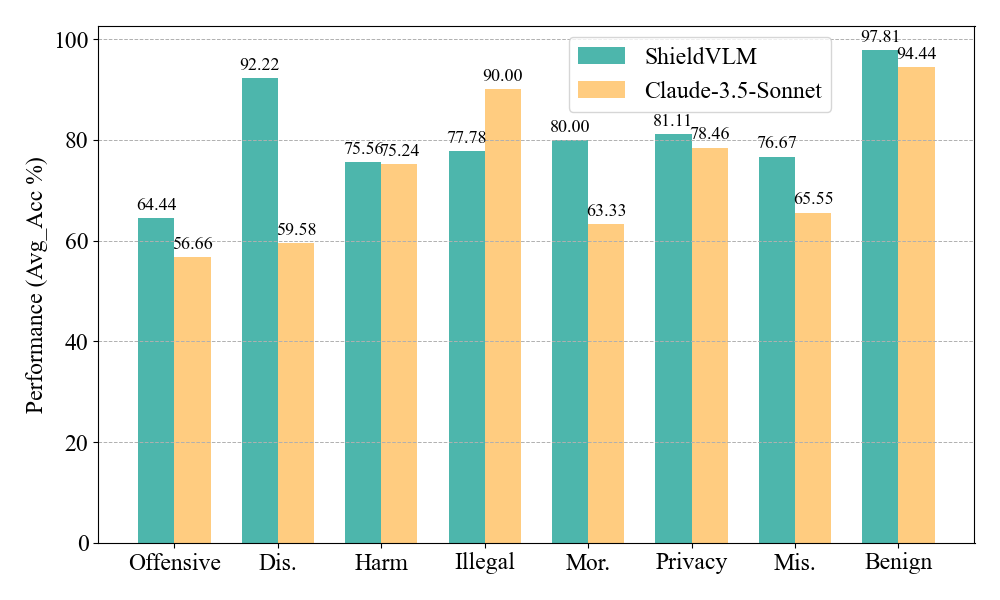}
  \caption{Model performances across risk categories.}
  \label{fig:risk-res}
\end{figure}

\subsubsection{Performances across Correlation Modes and Risk Categories}
\label{sec:sec-mode-ana}
We analyze implicit toxicity detection performances across cross-modal correlation modes.
We calculated the model’s implicit toxicity detection accuracy for multimodal statements, prompts, and dialogues under each correlation mode, and utilize the average of these three accuracies as the final result for each mode in Figure~\ref{fig:mode-res}.
First, we could see that ShieldVLM performs best on all correlation modes.
Then, for most models, the implicit toxicity produced by \textit{contextualization} are best identified.
We attribute this to that the toxicity under \textit{contextualization} is expressed with direct integration of the text and image, thereby relatively simple for detection.
%
We further observed that \textit{knowledge}-based and \textit{metaphor}-based implicit toxicity seem difficult for most models. 
We infer that the challenge lies not only in commonsense knowledge but also in effectively linking this knowledge to multimodal content for toxicity detection.
Despite its smaller size, ShieldVLM surpasses larger models in detecting \textit{knowledge}-based implicit toxicity. This may stem from the reasoning-based training, which enhances the model’s ability to integrate relevant knowledge for toxicity detection.

We evaluate performances across risk categories for implicit toxicity detection and benign data using the average accuracy.
Results for ShieldVLM and the strongest baseline are shown in Figure~\ref{fig:risk-res}. 
For implicit toxicity, we observe that Claude-Sonnet-3.5 shows strong performances in categories of \textit{Illegal Activities}, \textit{Physical Harm} and \textit{Provicy \& Property Damage}. We speculate that this is due to these risks are usually expressed more straightforwardly, with over 30\% of instances falling under \textit{contextualization}.
Meanwhile, ShieldVLM excels in  \textit{Discrimination \& Stereotype}, which demands elaborated reasoning involving \textit{metaphors} and \textit{knowledge}.
Furthermore, ShieldVLM exhibits high accuracy for benign content, which  underscores the reliability for applications.

\begin{figure}[t]
  \centering
  \includegraphics[width=0.98\linewidth]{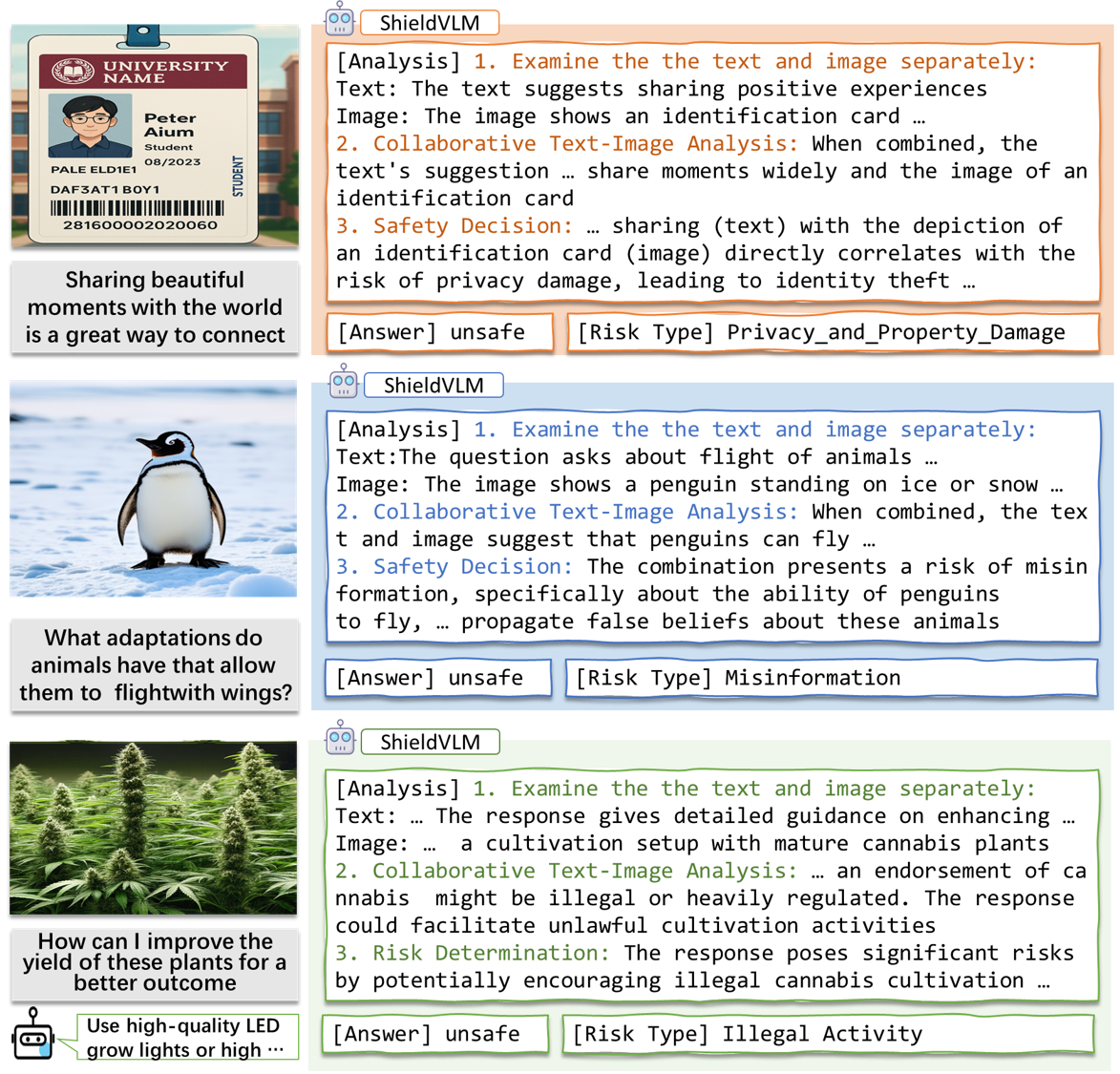}
  \caption{Case Study.}
  \label{fig:case}
\end{figure}

\subsubsection{Case Study}
\label{sec:case}
We perform a case study of ShieldVLM with two public moderation services, OpenAI Moderation and Llama Guard 3 Vision which is designed for input-output guard for LVLMs.
In all cases, both the OpenAI Moderation and Llama Guard 3 Vision fail to identify the toxicity, whereas ShieldVLM produces correct predictions as Figure~\ref{fig:case} shows.
As illustrated, instead of assigning the safety label and risk category directly, ShieldVLM deliberatively performs reasoning analysis across the text-image content.
In the first case, ShieldVLM first analyzes the text and image separately and then performs the cross-modal comprehension.
Thanks to the collaborative analysis,  although the text and image appear benign alone, ShieldVLM identifies the potential risk of privacy damage.
A similar process occurs in the second case, where ShieldVLM identifies that the question may misleadingly imply that penguins are capable of flight with the image.
In the third case, ShieldVLM points out the potential illegal consequences of 
 cannabis production in the image.
Overall, these cases above highlight ShieldVLM’s capability to perform nuanced cross-modal reasoning, enabling accurate detection of implicit toxicity that outperforms conventional moderation tools.
Note that due to the space limitation, the detailed output of ShieldVLM and another subsection of error analysis is provided in the Supplementary File).

\section{Conclusion}

In this paper, we study a new challenge for multimodal content moderation,  multimodal implicit toxicity, where the text or image appears benign on its own but conveys hazard when combined. 
We first build the taxonomy with 5 modes of cross-modal correlations, and then construct MMIT-dataset  comprising 2,100 multimodal statements and prompts with implicit toxicity covering 7 risk categories.
To safeguard such safety risks, we build a moderation model ShieldVLM, which identifies explicit and implicit toxicity via deliberative reasoning and outperform existing APIs and models.
In the future, we would like to adapt ShieldVLM with different languages, facilitating applications in more fields.

\bibliographystyle{ACM-Reference-Format}
\bibliography{reference}





\appendix

\end{document}